\documentclass[epjCONF]{svjour}
\usepackage{graphicx}
\graphicspath{{figures/}}

\usepackage[varg]{txfonts} 
\usepackage[latin1]{inputenc}
\session-title{%
19$^{\textnormal{\footnotesize th}}$ International %
  IUPAP Conference on Few-Body Problems in Physics%
}

\def\kaos{{\sc Kaos}\@}
\begin{document}
\title{%
  Associated {\boldmath $\Lambda/\Sigma^0$} electroproduction with the
  Kaos spectrometer at {MAMI} }%
\author{%
  P. Achenbach\thanks{\email{patrick@kph.uni-mainz.de}} on behalf of
  the A1 Collaboration}
\institute{%
  Institut f\"ur Kernphysik, Johannes Gutenberg-Universit\"at,
  J.-J.-Becherweg 45, D-55099 Mainz.\\
}
\abstract{ An instrument of central importance for the strangeness
  photo- and electroproduction at the 1.5\,GeV electron beam of the
  MAMI accelerator at the Institut f\"ur Kernphysik in Mainz, Germany,
  is the newly installed magnetic spectrometer \kaos\ that is operated
  by the A1 collaboration in $(e,e'K)$ reactions on the proton or
  light nuclei. Its compact design and its capability to detect
  negative and positive charged particles simultaneously complements
  the existing spectrometers. The strangeness program performed with
  \kaos\ in 2008--9 is addressing some important issues in the field
  of elementary kaon photo- and electroproduction reactions. Although
  recent measurements have been performed at Jefferson Lab, there are
  still a number of open problems in the interpretation of the data
  and the description of the elementary process using phenomenological
  models. With the identification of $\Lambda$ and $\Sigma^0$ hyperons
  in the missing mass spectra from kaon production off a liquid
  hydrogen target it is demonstrated that the extended facility at
  MAMI is capable to perform strangeness electroproduction
  spectroscopy at low momentum transfers $Q^2 < 0.5$\,(GeV$/c)^2$. The
  covered kinematics and systematic uncertainties in the cross-section
  extraction from the data are discussed. }
\maketitle
%
\section{Introduction}
%
Analogous to the successful description of pion photoproduction in the
$\Delta$ resonance region or $\eta$ photoproduction in the second
resonance region, phenomenological models can describe the
electromagnetic kaon production amplitudes associated to
$\Lambda/\Sigma^0$ hyperons. Many experimental data for kaon
production has been collected sofar by Jefferson Lab, ELSA, SPring8,
GRAAL, and Tohoku. Although recent measurements with high statistics
have been performed at Jefferson
Lab~\cite{Mohring2003,Carman2003,Ambrozewicz2007}, there are still a
number of open problems in the interpretation of the data and the
description of the elementary process using phenomenological models,
see Ref.~\cite{Mart2009} for a recent discussion.

Theoretical groups have developed a particular type of effective
Lagrangian model, commonly referred to as isobar approach, in which
the reaction amplitude is constructed from $s$-, $t$-, or $u$-channel
exchange diagrams.  Most of the models use single-channel approaches,
in which a single hadron is exchanged.  Since several resonances may
contribute in this channel, models disagree on their relative
importance, and many free parameters have to be fixed. Experimental
set-ups used to study the strangeness production channels are missing
forward angle acceptance leading to a strong variation of the models
in the forward region. Concerning the data published after 1990
discrepancies between the CLAS~\cite{Glander2003} and
SAPHIR~\cite{Bradford2005} experiments do not allow for a unique
description of the process. This situation clearly indicates that more
experimental and theoretical work is needed in order to provide a
comprehensive understanding of the elementary reaction.  A number of
new experiments are now addressing these issues, among them the
charged kaon electroproduction program with the \kaos\ spectrometer at
MAMI.

%
\section{Kinematics and experimental setup}
\begin{figure}
  \centering
  \includegraphics[width=\columnwidth]{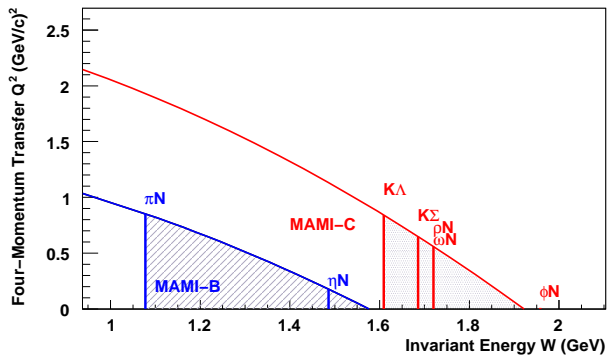}
  \caption{Kinematic regions accessible in electroproduction by the
    electron accelerator facility. The end-point energy of stage
    MAMI-B is 855\,MeV and of stage MAMI-C is 1508\,MeV (as of Summer
    2009).  The threshold energies for $\pi/\eta/\rho/\omega$ and
    $\phi$ production off the nucleon and the associated strangeness
    channels $K\Lambda$ and $K\Sigma$ are indicated; the kinematic
    regions for pion and kaon production are shaded. Only at MAMI-C
    the electroproduction of open strangeness is possible.}
  \label{AchenbachP_fig:MAMIelectroproduction}
\end{figure}

%
The accelerator facility MAMI at the Institut f\"ur Kernphysik in
Mainz, Germany, can now be used to study elementary strangeness
processes as the threshold energy was crossed by the 1508\,MeV
end-point energy of stage~C compared to the 855\,MeV end-point energy
of stage~B.  The kinematic regions accessible in electroproduction by
the electron accelerator facility are shown in
Fig.~\ref{AchenbachP_fig:MAMIelectroproduction}.

In a typical two-arm strangeness electroproduction experiment a
charged kaon is detected in coincidence with a scattered electron and
the recoiling hadronic system remains unobserved.  The incident
electron of energy $E$ scatters by radiating a virtual photon,
$q^\mu$. The scattered electron of energy $E'$ is emitted at a polar
angle $\theta$ with respect to the direction of the incident beam. The
virtual photon carries momentum, $\vec{q}$, and energy, $q_0$,
defining the momentum transfer squared $Q^2=-q^\mu q_\mu$.

The goals of the first measurements at MAMI were to determine the
angular dependence of the virtual photoproduction cross-section at
very low $Q^2$. This cross-section may be related to the five-fold
differential electroproduction cross-section via:
\begin{equation}
  \frac{d^{5}\sigma}{dE'd\Omega_{e} d\Omega_{K}^{*}} = \Gamma_v 
  \frac{d^{2}\sigma}{d\Omega_K^*}(W, Q^2, \epsilon, \theta_K^*, \phi_K^*)\,,
\end{equation}
where the kaon angles $\theta_K^*$ and $\phi_K^*$ are given in
spherical coordinates in the hadronic center-of-mass system and the
flux factor of virtual photons per scattered electron into $d E' d
\Omega_{e}$ is given by:
\begin{equation}
  \Gamma_v = \frac{\alpha}{2\pi^2} \frac{E'}{E} 
  \frac{\left|\vec{q}\right|}{Q^2} \frac{1}{1-\epsilon}\,,
\end{equation}
and the transverse polarization factor of the virtual photon is
denoted by $\epsilon$:
\begin{equation}
  \epsilon = \big( 1+2\frac{{\vert \vec{q}\vert}^2}{Q^2}{\tan^2
    \frac{\theta}{2}} \big)^{-1}\,.
\end{equation}

First strangeness production experiments were carried out at the
spectrometer facility of the A1 Collaboration~\cite{Blomqvist1998}.
An instrument of central importance for this program is the newly
installed magnetic spectrometer \kaos, dedicated to the detection of
charged kaons. Its compact design and its capability to detect
negative and positive charged particles simultaneously complements the
existing spectrometers.

\begin{figure}
  \centering
  \includegraphics[height=\columnwidth,angle=90]{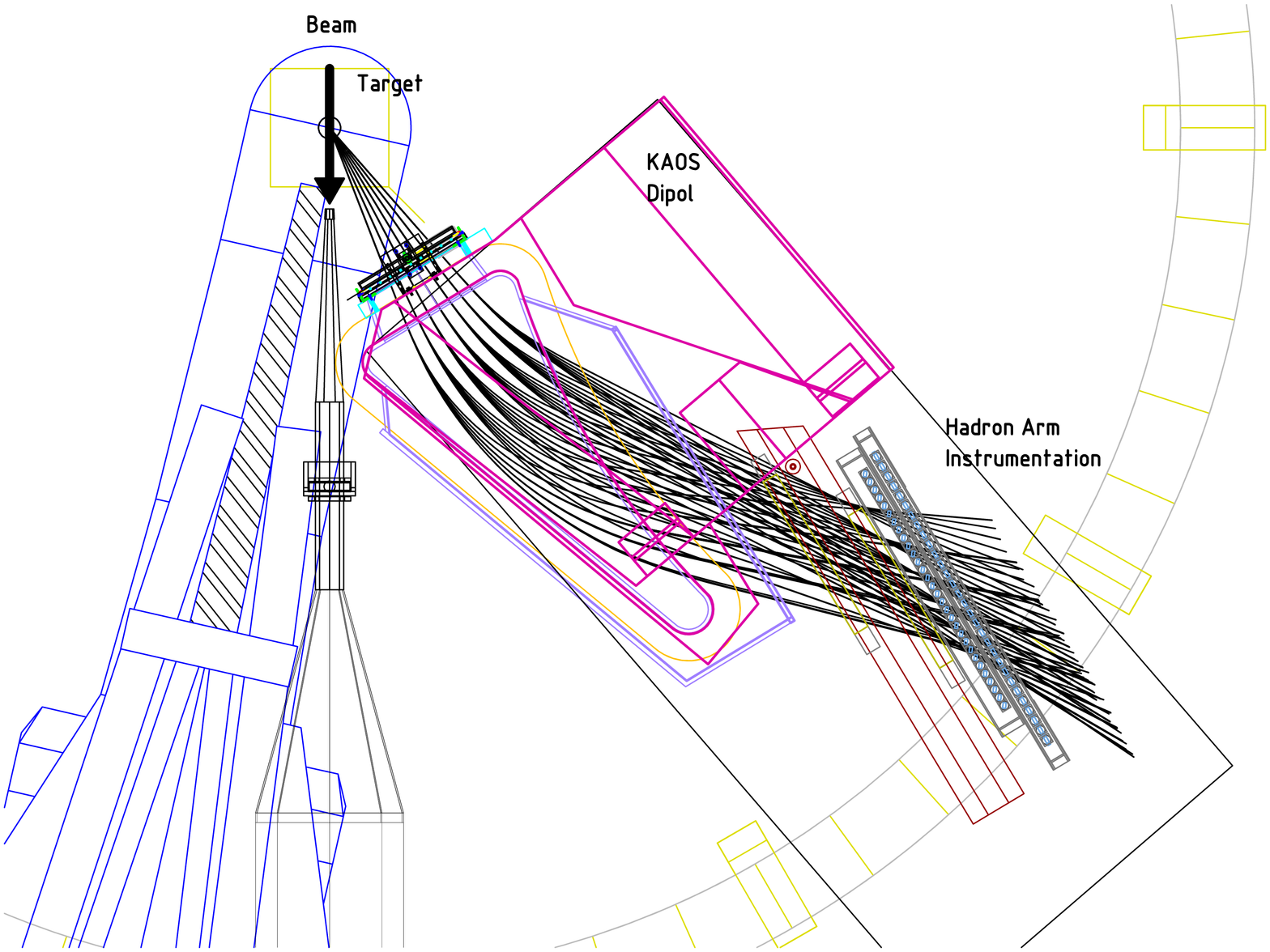}
  \caption{Overview of the first spectrometer set-up for strangeness
    electroproduction at the Mainz Microtron MAMI. Ray-traced particle
    trajectories through the \kaos\ spectrometer are shown by full
    lines.  The instrumentation of the hadron arm is indicated. Both
    spectrometers were positioned as close to the exit beam-line as
    mechanically possible.}
  \label{AchenbachP_fig:KAOS}
\end{figure}

Fig.~\ref{AchenbachP_fig:KAOS} shows a schematic drawing of the \kaos\
spectrometer as realized in the spectrometer hall. \kaos\ is operated
in a single-dipole configuration. The first-order focusing is achieved
with a bending of the central trajectory by $\sim$ 45$^\circ$ with a
momentum dispersion of 2.2\,cm$/$\%.  The dispersion to magnification
ratio of the \kaos\ hadron arm is $D/M_x$ = 1.2\,cm$/$\% which leads
to a first-order resolving power, $R_1 =$ $D/(M_x\sigma_x)$, of 2\,400
for a beam-spot size, $\sigma_0$, of 0.5\,mm.  The first order
momentum resolution, $\Delta p/p = \sqrt{(\sigma_x^2 +
  M_x^2\sigma_0^2)/D^2}$, is determined from these values to be of the
order of $\sim$ 10$^{-3}$, where the spatial resolution in the focal
plane, $\sigma_x$, was approximated with the anode wire distance of
2\,mm in the MWPC.

\begin{figure}
  \centering
  \includegraphics[width=\columnwidth]{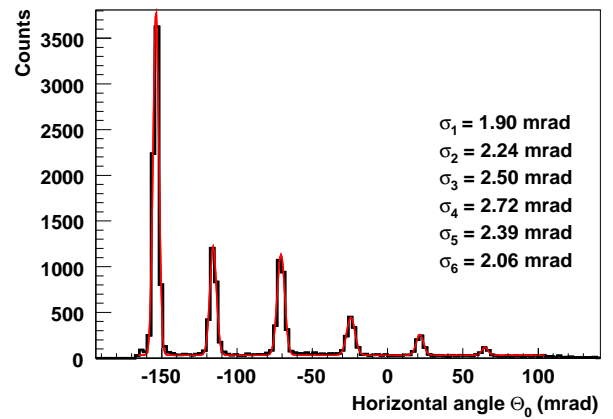}
  \caption{Preliminary analysis of elastically scattered electrons of
    450\,MeV energy, back-traced from the detector system of the
    \kaos\ spectrometer to the target, before applying any corrections
    of the aberrations. The sieve-slit collimator was made of a 20\,mm
    thick lead block with holes of angular width of $\sigma \sim$
    2.4\,mrad at 1000\,mm distance to the target. The peaks in the
    projection on the axis of the target angle $\theta_0$ in the
    dispersive plane were fitted by Gaussian distributions, resulting
    in widths of $\sigma \sim$ 1.9--2.7\,mrad.}
  \label{AchenbachP_fig:sieveslit}
\end{figure}

Detailed measurements with electron beams elastically scattered off
$^{12}$C and $^{181}$Ta foils have been performed to study the
properties of the magnet optics. To select well determined angles in
both planes a sieve collimator, made of 20\,mm thick lead with 16
holes, 5\,mm in diameter each, was installed in a distance of 1000\,mm
from the target in front of the magnet. For each hole the sieve
collimator defines primary angle distributions with a width of $\sigma
=$ 2.45\,mrad. Electrons passing different holes are clearly separated
in the reconstructed coordinates.  The deviations to the nominal hole
coordinates quantify the uncorrected optical aberrations of the
system. The magnitude of the necessary correction of the transfer
matrix elements is of the order of 10\,mrad or more, as can be deduced
from the deviations of the measured hole positions to the nominal
positions. The projections of the events to the target angle
$\theta_o$ in the dispersive plane produced peaked distributions width
widths of $\sigma \sim$ 1.9--2.7\,mrad, as shown in
Fig.~\ref{AchenbachP_fig:sieveslit}.  A finer collimator was prepared
in 2009 for an improved determination of the transfer matrix
elements. It contained 53 holes in 5 different horizontal lines along
the dispersive plane. The calibrations are ongoing and no final
results from the experiments can be provided unless the calibration
runs are fully analyzed.

In 2008 and 2009 strangeness production off the proton was measured at
two different kinematic settings with this experimental set-up. The
beam from MAMI-C impinged with an energy of 1507\,MeV on a
liquid-hydrogen target.  Positive kaons were detected in the \kaos\
spectrometer benefiting from the large in-plane angular acceptance at
scattering angles $\theta_K=$ 21--43$^\circ$, and of the large
momentum acceptance at $p =$ 400--700\,MeV$/c$.  The scattered
electrons from the H$(e,e'K)$ reaction were detected in the
high-resolution magnetic spectrometer B (SpekB) that was kept fixed at
an angle of $\theta \approx$ -15$^\circ$ during the experiments.

\begin{table}
  \caption{Kinematics of the 2009 beam-time on kaon elementary 
    electroproduction at the Mainz Microtron.}
\label{AchenbachP_tab:kinematics}
\begin{tabular}{lccc}
\hline\noalign{\smallskip}
     initial electron energy, $E$ (MeV) &                     1508\\
     final electron energy, $E'$  (MeV) &                      327\\
     electron scattering angle, $\theta$ (deg) &              15.5\\
     degree of transverse polarization, $\epsilon$ &         0.406\\
     total energy in $cm$ system, $W$ (MeV) &                 1745\\
     virtual photon momentum transfer, $Q^2$ (GeV$^2/c^2$) &   .036\\ 
     virtual photon flux factor, $\Gamma_v$ (ph./GeV/sr) &    .004\\ 
\noalign{\smallskip}\hline
\end{tabular}
\end{table}

The luminosity was calculated from the density and thickness of the
target, and the beam current, and is corrected for the DAQ dead-time.
The collected data in 2008 corresponds to effectively 3 days of
beam-time at 1--4\,$\mu$A beam current for each kinematic point with a
total integrated luminosity of $\int\! {\cal L} d t \sim$
284\,fbarn$^{-1}$.  One setting with a central momentum in \kaos\ of
530\,MeV$/c$ was continued by a longer data-taking campaign on kaon
elementary electroproduction in 2009 for a statistically and
systematically improved data-set. In
Table~\ref{AchenbachP_tab:kinematics} a summary on this particular
kinematics is given.  The current was held constant at 2, respectively
4\,$\mu$A.  With a run-time of 265\,h using 2\,$\mu$A beam current (at
13\,\% dead-time) and 40\,h using 4\,$\mu$A beam current (at 44\,\%
dead-time) the accumulated and corrected luminosity for the 2009
data-taking campaign was $\int\! {\cal L} d t \sim$
3000\,fbarn$^{-1}$.

\section{Particle tracking and identification with Kaos}
%
The tracking of particles through \kaos\ is performed by means of two
large MWPC with a total of 2 $\times$ 310 analogue channels.  Five
cathode wires are connected together and are brought to one charge
sensitive preamplifier followed by an ADC card.  The transputer-based
read-out system is connected to a multi-link card of a front-end
computer. To determine the particle track the measured charge
distributions are analyzed by the center-of-gravity method.

In the subsequent analysis of the data not only the particle track is
used, but a large set of track quality factors.  For example, the
correlation that exists between the induced charge measured in one
cathode plane to the induced charge of the perpendicular plane allows
in many cases a correct pairing of clusters. Valid tracks are bound to
angular limits, given by the acceptance of the
spectrometer. Especially the relation between vertical hit positions
of the two chambers provides a powerful criterion for the track
finding. The multi-hit capability of the MWPC is important to allow
for beam currents above 1\,$\mu$A. As long as two tracks are
sufficiently separated in space to induce distinct charge
distributions on both cathode planes, the double track ambiguity can
be resolved with high efficiency.

In 2009 two small scintillating detectors of dimensions $\sim 2 \times
3$\,cm$^2$ with 5\,mm thickness were installed in front of each MWPC
to determine the tracking and detection efficiency of the hadron arm.
The detectors can easily be moved out of the spectrometer's acceptance
when not in use.

Particle identification in the \kaos\ spectrometer is based on the
particle's time-of-flight and its specific energy loss. A segmented
scintillator wall with 30 paddles read out at both ends by fast
photomultipliers is located near the focal-surface and measures the
arrival time. A second wall with 30 paddles is used to discriminate
valid tracks against background events. A top--bottom mean timing for
deriving the trigger is performed by summing the analogue signals. The
signal amplitudes were corrected for the particle's pathlength
through the scintillator bulk material and the light absorption inside
the paddle.  Because of aging of the material the absorption of the
scintillation light inside the scintillator is strong.

The time spectrum is systematically broadened by the propagation time
dispersion inside the scintillator, the time differences between
different scintillator paddles and their associated electronic
channels, and by the variation of the time-of-flight, being
proportional to the pathlength through the spectrometer.  The
Gaussian width of the $(e',\pi)$ peak after corrections is $\Delta
t_{\it FWHM} =$ 1.07\,ns, which is a typical inter-spectrometer time
resolution.

From the point of view of background events in the detectors the
set-up of the \kaos\ spectrometer at forward angles is
problematic. Shielding walls were installed comprising neutron
shielding material (borated polyethylene) and lead for the
electromagnetic shielding. In order to cope with the remaining high
background rates \kaos\ has been equipped with a modern trigger
system.  The correlation between the paddles in both walls
corresponding to valid trajectories was employed for the trigger
decision.  The relatively high complexity of the valid patterns and
the intrinsic necessity of a flexible trigger logic for a spectrometer
that faces a wide experimental program demanded a programmable logic
system which was realized in so-called VUPROM modules. Each so called
VME Universal Processing Module (VUPROM) has 256 I$/$O channels in a
1-unit-wide VME 6U form-factor.  The digital signals are transmitted
in high speed differential LVDS standard and are combined in 8 groups
of 32 channels each, where one group is reserved for output, three
groups are allocated to input, and the remaining four groups are
freely configurable as input or output.  Such a module is equipped
with one {\sc Virtex-4} FPGA, capable of operation at 400\,MHz, and
one 1\,GHz DSP.  The coincidence trigger is based on time-of-flight
and is therefore generated from the VUPROM logic and from SpekB
scintillator signals.
    
Some more selected properties of the optics of the \kaos\
spectrometer, its kaon identification capabilities, and existing or
planned modifications to operate \kaos\ as a double-arm spectrometer
under zero degree scattering angle can be found in
Ref.~\cite{Achenbach2009:Sendai}.

\section{Data analysis}
%
%
\begin{figure}
  \centering
  \includegraphics[height=\columnwidth,angle=-90]{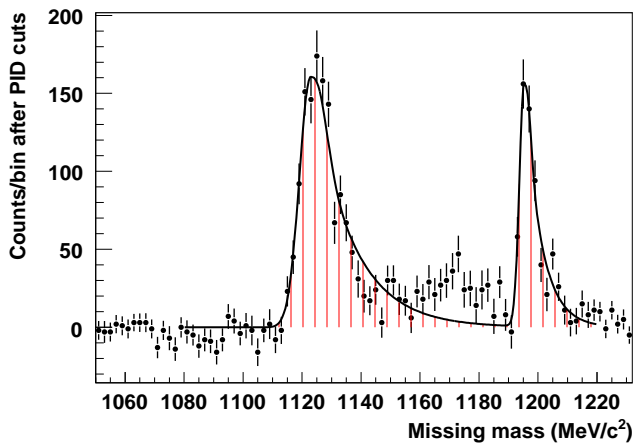}
  \caption{Preliminary missing mass spectrum in the H$(e,e'K^+)$
    reaction from the 2009 data-taking campaign as discussed in the
    text. The mass resolution is limited by the errors in the
    estimated transfer matrix that was not yet corrected for the
    aberrations seen in calibration measurements. The broad structure
    at 1170\,MeV$/c^2$ is believed to be caused by particles scattered
    inside the spectrometer. The background in two averaged $(e',K)$
    coincidence time side-bands was subtracted with the appropriate
    weight.}
  \label{AchenbachP_fig:MissingMass}
\end{figure}

Kaon identification is based on specific energy-loss and
time-of-flight, electron identification on a signal in a gas Cherenkov
counter.  The flight time difference between protons and kaons is
10--15\,ns, between pions and kaons 5--10\,ns. The energy-loss
separation between kaons and pions is small, whereas the separation
between kaons and protons is of the order of 2--5\,MeV$/$cm.  The
following cuts were applied to a limited data set taken with 2\,$\mu$A
beam current in 2009:

\begin{description}
\item[Electron identification:] Finite sum of Cherenkov counter
  amplitudes to reject pions in the electron arm.
\item[Spectrometer fiducial volume:] $\theta_0 >$ -10.5\,$^\circ$ and
  $\theta_0 < 8^\circ$ and $\theta_0 < 0.475 \times (\Delta p +
  20\,\%)$ to restrict the acceptance to a region where agreement
  between the Monte Carlo code and the analyzed data was excellent.
\item[MWPC track reconstruction:] MWPC quality $>$ 0.0001 ensuring
  good quality tracks in the hadron arm.
\item[Particle identification:] Positive squared missing mass for a
  proper missing mass reconstruction, and $p_\mathit{Kaos} <$ 1.8
  $\times$ $p_\mathit{central}$ for a proper momentum reconstruction,
  and $\beta_\mathit{Kaos} >$ 0.4 to reject erroneous timing values.
\item[Kaon identification:] $|t^K_{F,G}| < 1.8$\,ns, being the
  measured TOF corrected for the expected kaon flight time, and
  $|\Delta E^K_{F,G}| < 400$\,keV, being the measured specific energy
  loss corrected for the expected kaon energy loss. A small $t^K$
  indicates that the correct mass hypothesis has been made while a
  small $\Delta E^K$ reduces the random background events.
\item[Reaction channel identification:] For the $\Lambda$ and
  $\Sigma^0$ hyperons 1110\,MeV$\!/c^2 < M_X <$ 1140\,MeV$\!/c^2$,
  respectively 1190\,MeV$\!/c^2 < M_X <$ 1205\,MeV$\!/c^2$.
\end{description}
\begin{figure}
  \centering
  \includegraphics[width=\columnwidth]{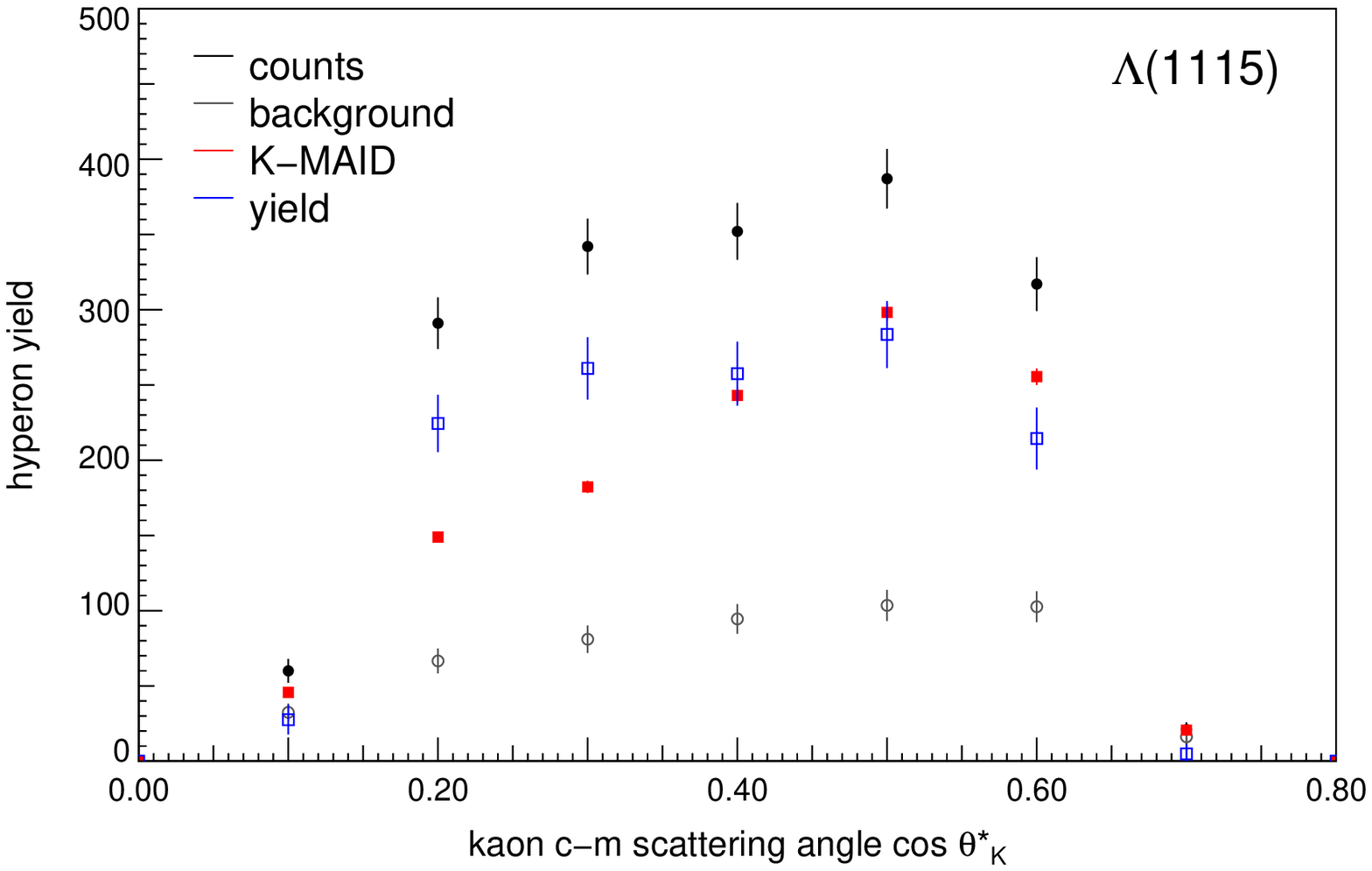}
  \includegraphics[width=\columnwidth]{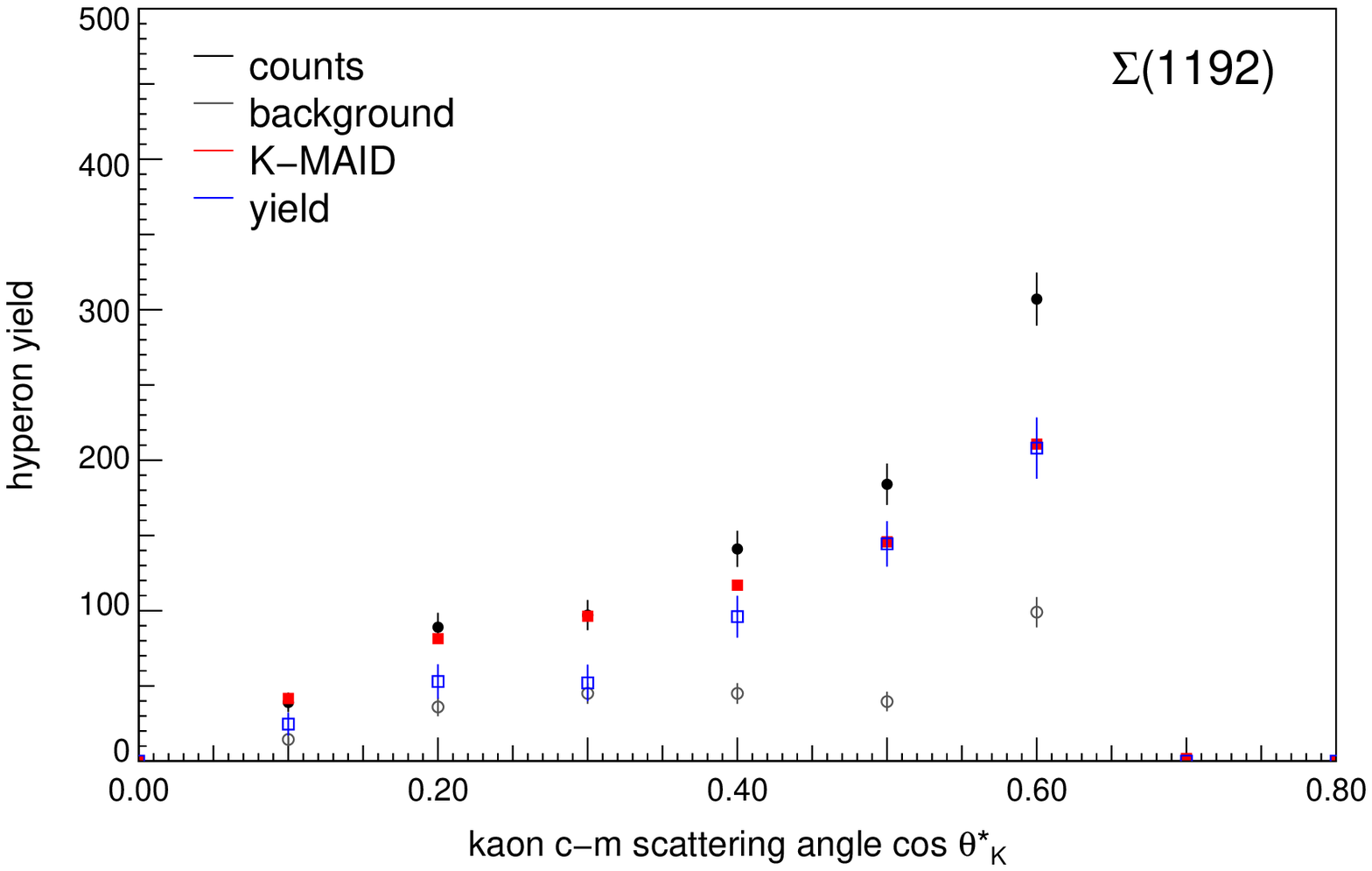}
  \caption{Preliminary yield of identified
    H$(e,e'K^+)\Lambda,\Sigma^0$ reactions as a function of the
    center-of-mass kaon scattering angle. The measured events (open
    round symbols) are corrected for background events (black round
    symbols) to obtain the hyperon yield (blue round symbols). The
    data is compared to a Monte Carlo simulation using the K-Maid
    model~\cite{KMAID} as an input for the cross-sections and a
    generalized acceptance function for the detector description. The
    background subtraction is still dependent on the calibration.}
  \label{AchenbachP_fig:KYield}
\end{figure}

After electron and kaon identification the measured momenta, in
magnitude and direction with respect to the incoming beam, allow for a
full reconstruction of the missing energy and missing momentum of the
recoiling system.  After application of all cuts on the data, the
remaining events are used to extract the kaon yield in the two hyperon
channels.  A cut was placed on the missing mass in order to separate
the $\Lambda$ and $\Sigma^0$ reaction channels. Because the $\Sigma^0$
peak resides on the tail of the $\Lambda$ peak, the latter has to be
accounted for in the analysis of the $\Sigma^0$-region of the measured
spectrum.  A preliminary missing mass distribution is shown in
Fig.~\ref{AchenbachP_fig:MissingMass}. The mass resolution is limited
by the errors in the estimated transfer matrix that was not yet
corrected by the aberrations seen in calibration measurements. Since
the \kaos\ spectrometer is operated as a single dipole with open yoke
geometry a significant fraction of detected particles have scattered
inside the spectrometer and need to be treated accordingly.

Random background events in the missing mass spectra were subtracted
by two averaged $(e',K)$ coincidence time side-bands with the
appropriate weights. The preliminary yields of identified
H$(e,e'K^+)\Lambda,\Sigma^0$ reactions as a function of the
center-of-mass kaon scattering angle are shown in
Fig.~\ref{AchenbachP_fig:KYield}. The background subtraction is still
dependent on the calibration. The data is compared to a Monte Carlo
simulation using the K-Maid model~\cite{KMAID} as an input for the
cross-sections and a generalized acceptance function for the detector
description.

The number of kaons that were actually detected was less than the true
number of kaons produced in the reaction, and corrections for their
decay had to be taken into account. The kaon survival fraction varied
between 0.2 and 0.35 for the range of momenta detected.

\section{Extraction of cross-sections}
%
The further analysis required a detailed Monte Carlo simulation of the
experiment in order to extract cross-section information from the
data.  The experimental yield can be related to the cross-section by
\begin{equation}
  Y = {\cal L} \times \int\!\! \Gamma(Q^2,W) 
  \frac{d^{2}\sigma}{d\Omega_K^{*}} A(d^{5}V) R(d^{5}V)\, 
  dQ^{2} dW d\phi_{e} d\Omega_{K}^{*}\,,
\end{equation}
where ${\cal L}$ is the experimental luminosity including global
efficiencies such as dead-times, $A$ is the acceptance function of the
coincidence spectrometer set-up including momentum dependent
corrections such as the tracking efficiency, and $R$ is the correction
due to radiative or energy losses.

\begin{figure}
  \centering
  \includegraphics[width=\columnwidth]{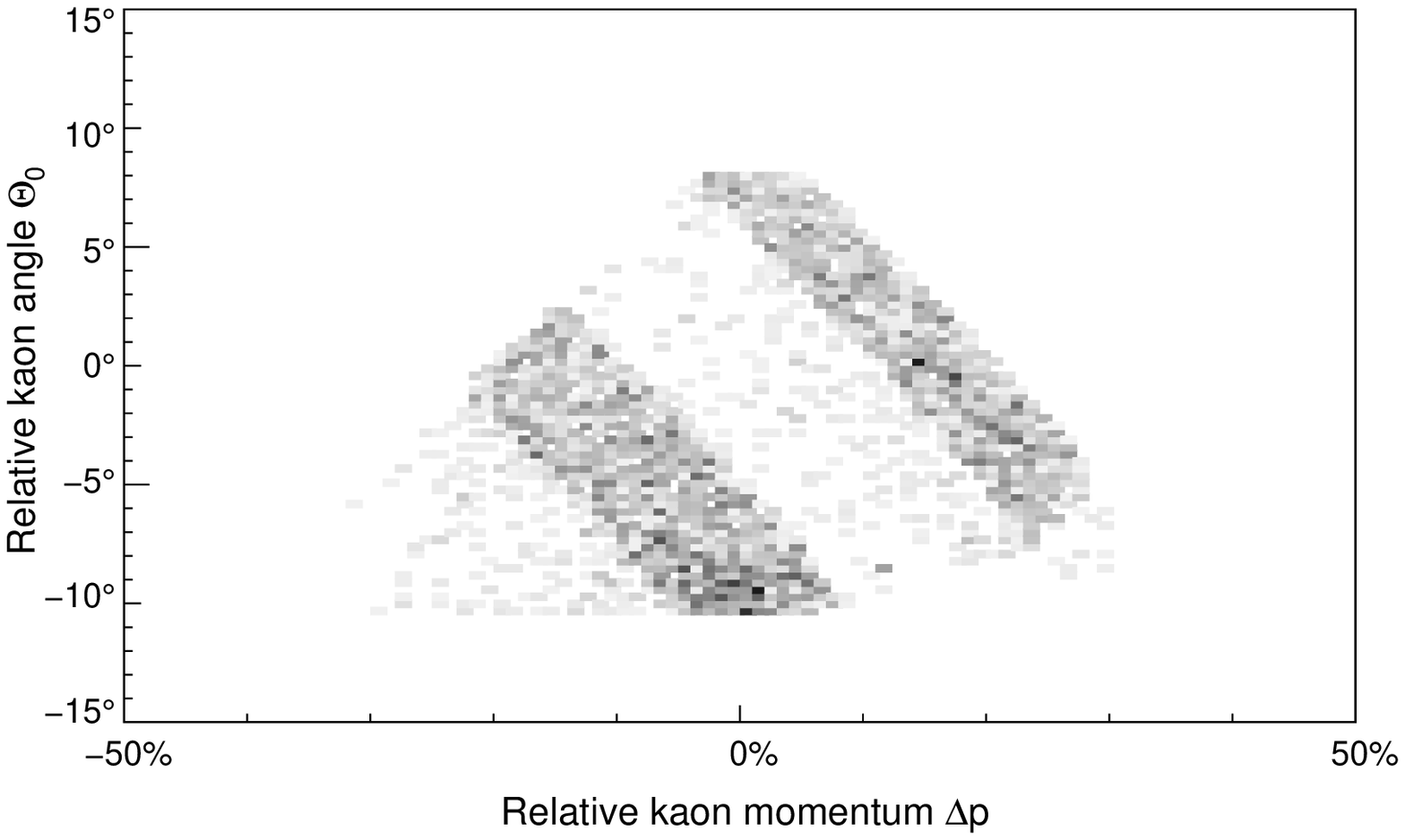}
  \includegraphics[width=\columnwidth]{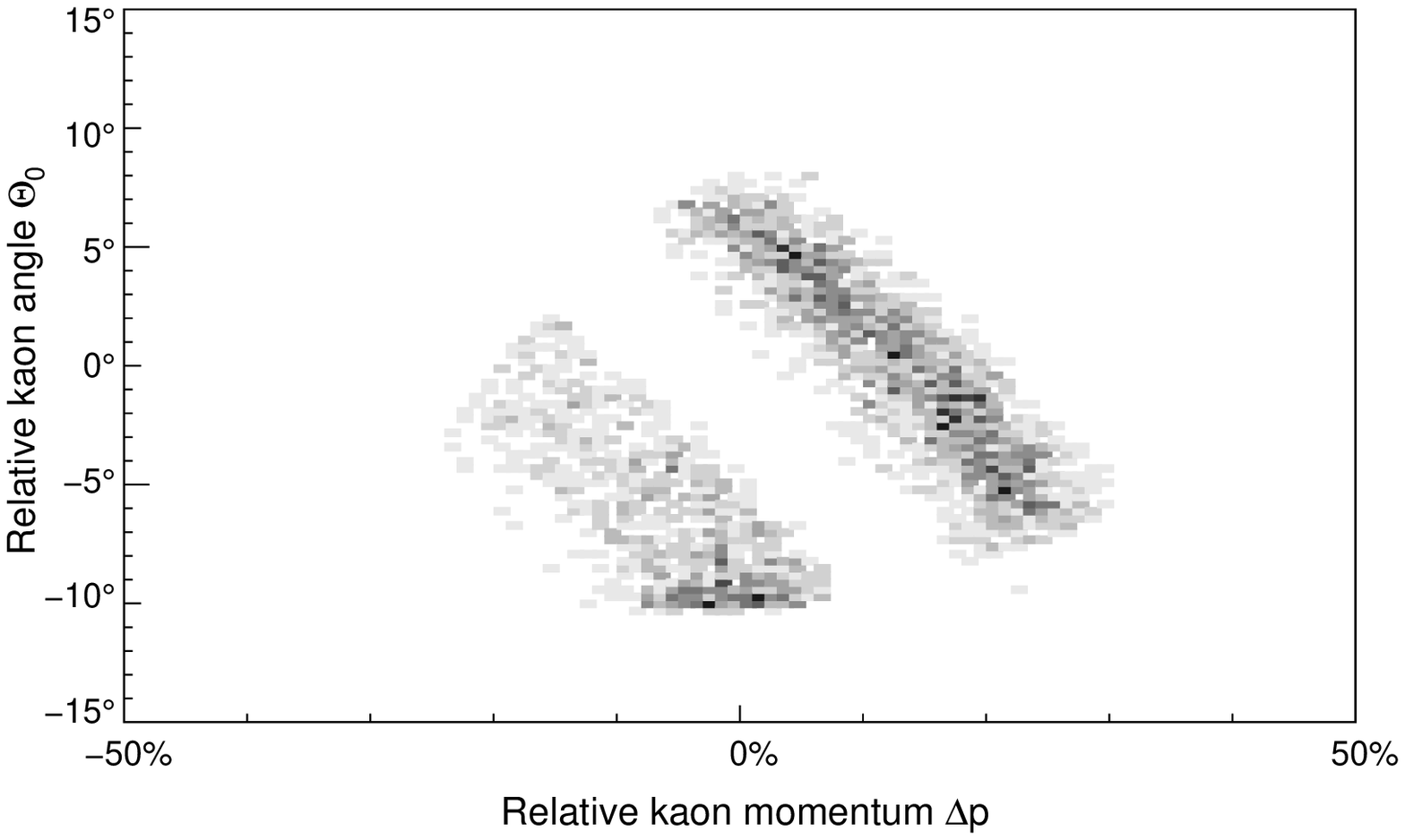}
  \caption{Phase-space in the target angle vs.\ momentum plane for
    kaon acceptance in the \kaos\ spectrometer simulated by a {\sf
      Geant4} Monte Carlo integration (top) and measured during the
    2009 beam-time (bottom). The kaon data was extracted after
    particle identification and missing mass cuts. The simulation
    includes radiative corrections and energy-loss in the target.}
  \label{AchenbachP_fig:phasespace}
\end{figure}

In the Monte Carlo for the integral is evaluated in the phase-space
$\Delta V = \Delta Q^2 \Delta W \Delta \phi_e \Delta \Omega_K^*$ with
limits that extended beyond the physical acceptances of the
spectrometers.  The simulation is giving the phase-space accepted by
the spectrometers according to a chosen kinematics together with the
radiative corrections, absorption and decay losses as well as detector
inefficiencies. By virtue of the Monte Carlo technique used to perform
this integral the quantities $A$ and $R$ are not available separately
or on an event-by-event basis. In Fig.~\ref{AchenbachP_fig:phasespace}
the phase-space in the target angle vs.\ momentum plane for kaon
acceptance in the \kaos\ spectrometer simulated by a Monte Carlo
integration method is compared to the data measured during the 2009
beam-time. The kaon data was extracted after particle identification
and missing mass cuts. The agreement inside the fiducial volume is
excellent.

The general acceptance can be written in terms of the cross-section at
a given point $({d^{2}\sigma}/{d\Omega_K^*})_{0,\theta_K^*}$, called
the scaling point, at $\langle Q^2 \rangle$, $\langle W \rangle$,
$\langle \phi_e \rangle$, and $\langle \phi_K^* \rangle$, when
studying the dependence on the remaining variable $\cos\theta_K^*$:
\begin{eqnarray}
  Y & = & {\cal L} \times 
  \left( \frac{d^{2}\sigma}{d\Omega_K^*} \right)_{0,\theta_K^*} \times \\
  & & \int\!\! \Gamma(Q^2,W) \frac{\frac{d^{2}\sigma}{d\Omega_K^*}}{ 
    \left( \frac{d^{2}\sigma}{d\Omega_K^*} \right)_{0,\theta_K^*}}
  A(d^{5}V) R(d^{5}V)\, 
  dQ^{2} dW d\phi_{e} d\Omega_{K}^{*} \nonumber
\end{eqnarray}
Then, the behavior of the cross-section across the acceptance needs to
be implemented according to a given theoretical description. We
applied the K-Maid~\cite{KMAID} isobar model, however, we could as
well compare to the full (SL) and the simplified version (SLA) of the
so-called Saclay-Lyon model~\cite{SLA}. The scaling point for the data
set from 2009 is at $\langle Q^2 \rangle =$ 0.036\,(GeV$\!/c)^2$,
$\langle W \rangle =$ 1.750\,GeV, $\langle \epsilon \rangle= $ 0.4.
Finally, the cross-section is evaluated dividing the yield by the
luminosity and the evaluated integral. An important observable to be
investigated is the stability of the results as a function of the
applied cuts and its influence on the extracted cross-section. Up to
now, systematic uncertainties being discussed in the next section
hinder the extraction of final values.

\section{Systematic uncertainties}
%
%
\begin{figure}
  \centering
  \includegraphics[width=\columnwidth]{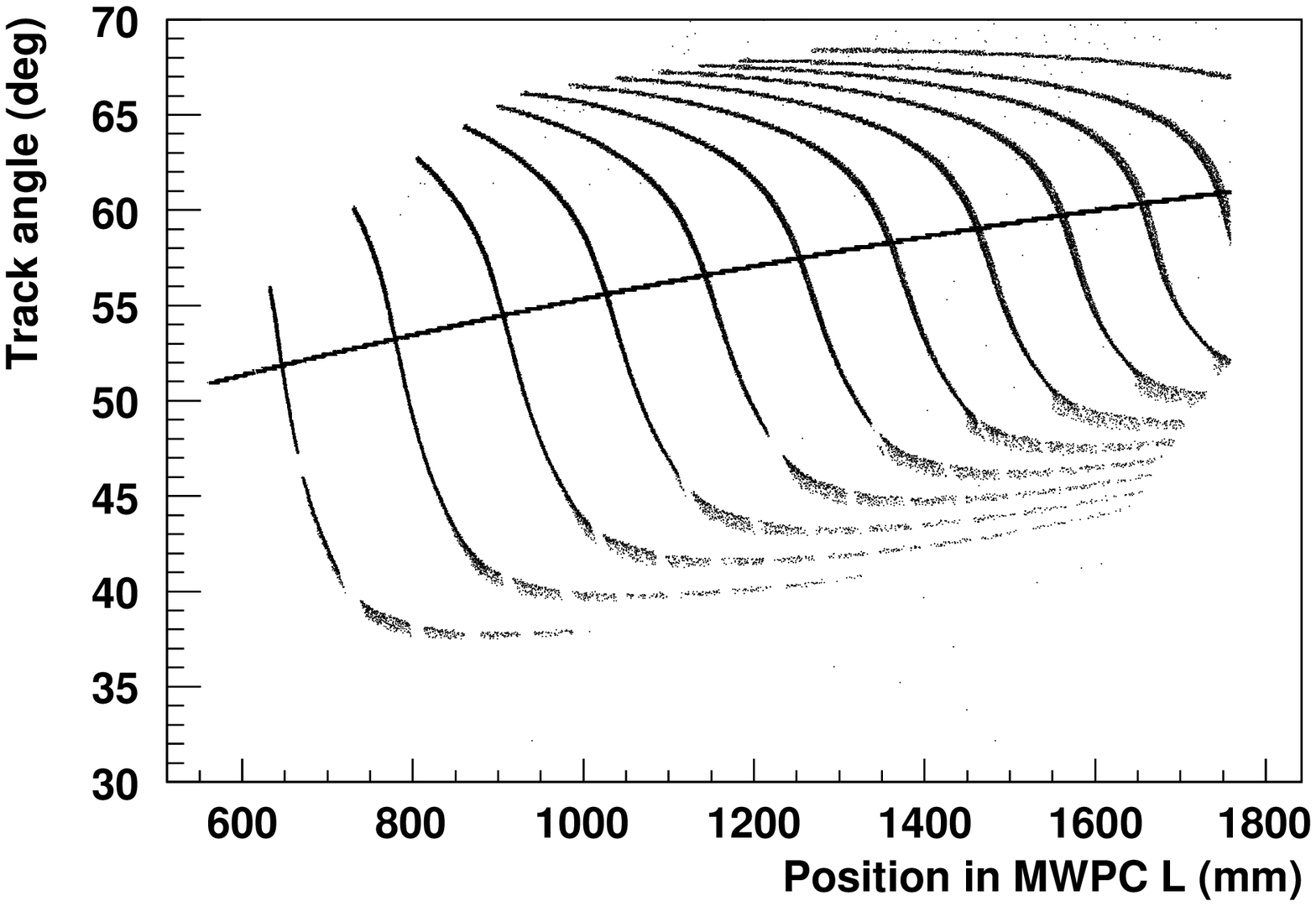}
  \includegraphics[width=\columnwidth]{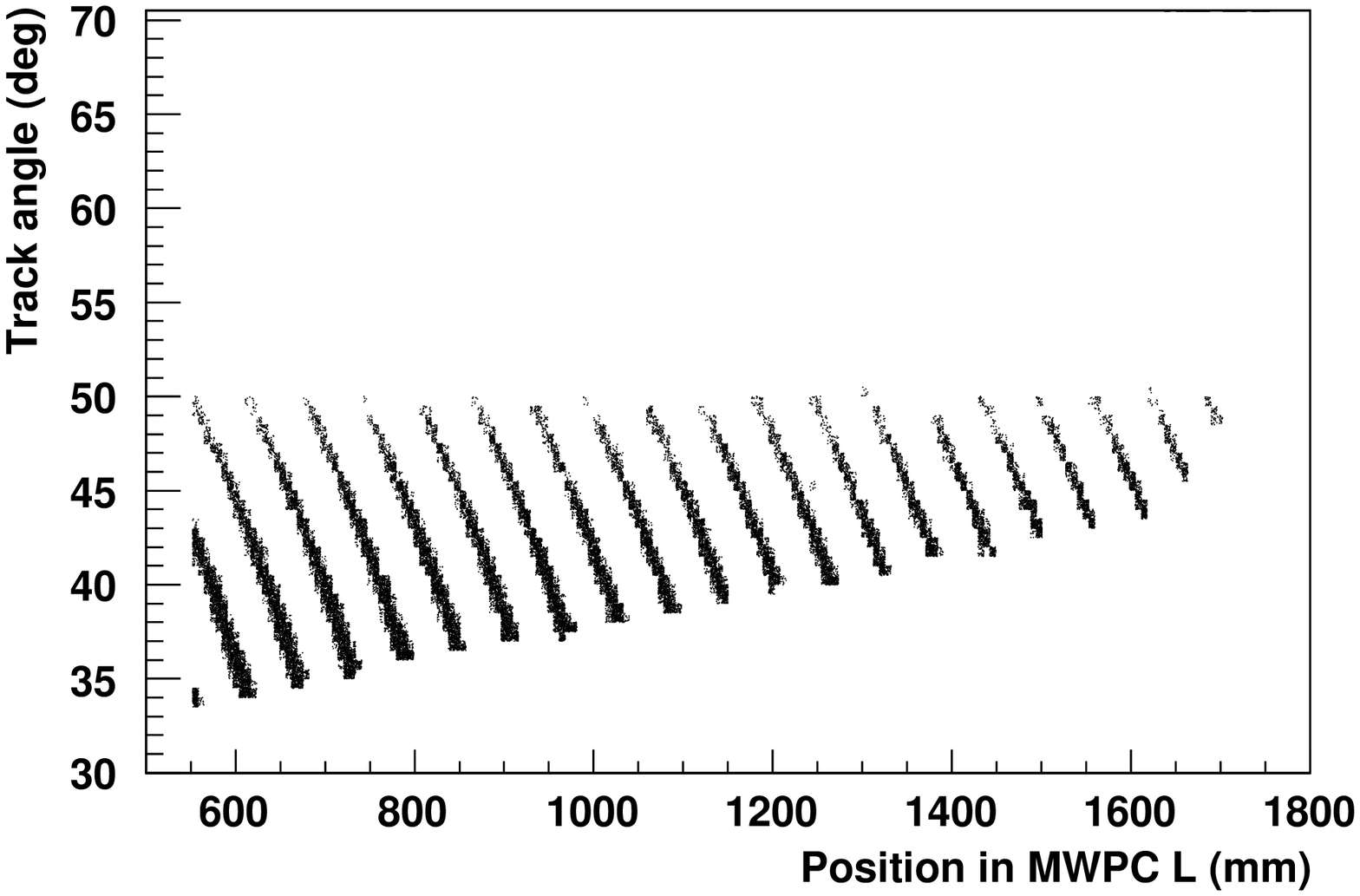}
  \caption{Top: Projection of simulated tracks for particles with
    momenta of 350\,MeV up to 600\,MeV in steps of 25\,MeV, over the
    full angular acceptance, and for particles of central target
    angles over the full momentum acceptance, in the coordinate system
    of the tracking chambers. In the regions with strong curvature the
    momentum depends on many higher-order terms of the transfer
    matrix. Bottom: Simulated geometrical scintillator wall
    inefficiencies in the same coordinate system. Both effects
    contribute to systematic uncertainties in the cross-section
    extraction.}
  \label{AchenbachP_fig:systematics}
\end{figure}

The total efficiency for the detection of kaons with the \kaos\
spectrometer can be factored into the intrinsic wire chamber (MWPC)
detection efficiency, the analysis cut efficiency, the trigger
efficiency, and the tracking efficiency.  In order to investigate the
MWPC, tracking, and trigger efficiencies dedicated measurements have
been performed in 2008 and 2009. The efficiency of the MWPC for the
identification of minimum ionizing particles is measured to be better
98\,\% with only small variations over time.  On the contrast, the
tracking efficiency is dependent on the hit multiplicities in the wire
chambers leading to ghost tracks that are classified according to a
likelihood method and that are matched to scintillator hits.  The
track reconstruction in the wire chambers represents a primary source
of inefficiency for beam currents of several $\mu$A.  For pions
tracking efficiencies ranged between 76 and 45\,\% corresponding to
beam currents of 1--4\,$\mu$A, for protons the efficiencies were
between 92 and 77\,\%.

In an ideal spectrometer the position along the focal surface at which
a particle is detected is directly related to its momentum. In
practice, however, due to kinematic broadening and spectrometer
aberrations this position will also depend on the angle of the
particle trajectory with the central
ray. Fig.~\ref{AchenbachP_fig:systematics}\,(top) shows the projection
of simulated tracks with momenta of 350\,MeV up to 600\,MeV over the
full angular acceptance in the coordinate system of the tracking
chambers.  Particles with given momenta but varying target angles
leave traces in this coordinate system that are lines of constant
slopes in the central region of the acceptance and are strongly curved
at the edges. The plot includes also the trace for particles of
central target angles over the full momentum acceptance. In the
regions with strong curvature the momentum depends on many
higher-order terms of the transfer matrix, which need to be precisely
determined.
    
The trigger was generated by a combination of hits in the two
scintillator walls.  Particular momentum and track angle combinations
allow the particle to cross the gaps between the scintillator paddles
giving rise to significant detector inefficiencies. In
Fig.~\ref{AchenbachP_fig:systematics}\,(bottom) a simulation of these
geometrical inefficiencies are shown in the tracking coordinate
system. Momentum and track angle dependent trigger efficiency
corrections need to be applied to the generalized acceptance.

\section{Summary}
%
Using the recently installed magnetic spectrometer \kaos\ and a
high-resolution spectrometer in coincidence it was possible to measure
elementary kaon electroproduction at MAMI in a kinematic regime sofar
not covered by Jefferson Lab experiments. At different settings we
have confirmed the particle identification power and tracking
capabilities of the \kaos\ spectrometer's hadron arm detection
system. The large background rates and inefficiencies encountered in
the 2008--9 data-taking campaigns are corrected using dedicated
calibration data and simulation procedures.  The systematic
uncertainties in the cross-section extraction are under study.


\begin{thebibliography}{99}
\bibitem{Mohring2003} R.~M. Mohring \textit{et al.} (E93-018
  Collaboration), Phys. Rev. \textbf{C 67}, (2003) 0552025.

\bibitem{Carman2003} D.~S. Carman \textit{et al.} (CLAS
  Collaboration), Phys. Rev. Lett. \textbf{90}, (2003) 131804.

\bibitem{Ambrozewicz2007} P.~Ambrozewicz \textit{et al.} (CLAS
  Collaboration), Phys. Rev. \textbf{C 75}, (2007) 045203.

\bibitem{Mart2009} T.~Mart, \textit{Proc. Intern. Symposium on
    Strangeness in Nuclear and Hadronic Systems (Sendai08)}
  (eds. K.~Maeda \textit{et al.}, Sendai, Japan, 2008), 51.

\bibitem{Glander2003} K.~H. Glander \textit{et al.} (SAPHIR
  Collaboration), Eur. Phys. J. \textbf{A 19}, (2004) 251.

\bibitem{Bradford2005} R.~Bradford \textit{et al.} (CLAS
  Collaboration), Phys. Rev. \textbf{C 73}, (2006) 035202.

\bibitem{Blomqvist1998} K. I. Blomqvist \textit{et al.} (A1
  Collaboration), Nucl. Instr. Meth.  Phys. Res. \textbf{A 403},
  (1998) 263.

\bibitem{Achenbach2009:Sendai} P.~Achenbach,
  \textit{Proc. Intern. Symposium on Strangeness in Nuclear and
    Hadronic Systems (Sendai08)} (eds. K.~Maeda \textit{et al.},
  Sendai, Japan, 2008), 332.
  
\bibitem{SLA} T. Mizutani, C. Fayard, G.-H. Lamot, and B. Saghai,
  Phys. Rev. \textbf{C 58}, (1998) 75.

\bibitem{KMAID} T. Mart and C. Bennhold, Phys. Rev. \textbf{C 61},
  (1999) 012201(R); http://www.kph.uni-mainz.de/MAID/
\end{thebibliography}
\end{document}